\documentclass[journal]{IEEEtran}
\pdfoutput=1

\usepackage{amsmath,amssymb,graphicx,epstopdf}

\newcommand{\acro}{ABLE}

\begin{document}

\title{Adaptive Ultrasound Beamforming using~Deep~Learning}

\author{Ben~Luijten,~\IEEEmembership{Student,~IEEE},
Regev Cohen~\IEEEmembership{Student,~IEEE},
Frederik J. de Bruijn,
Harold A.W. Schmeitz, \\
Massimo Mischi,~\IEEEmembership{Senior Member,~IEEE,} 
Yonina C. Eldar,~\IEEEmembership{Fellow,~IEEE,} \\and 
Ruud J.G. van Sloun,~\IEEEmembership{Member,~IEEE,}

\thanks{B. Luijten, M. Mischi and R.J.G. van Sloun are with the department of Electrical Engineering, Eindhoven University of Technology, Eindhoven, The Netherlands. (e-mails: w.m.b.luijten@tue.nl, m.mischi@tue.nl, r.j.g.v.sloun@tue.nl)}
\thanks{R. Cohen is with the department of Electrical Engineering, Techion Israel Institute of Technology, Haifa, Israel. (e-mail: regev.cohen@gmail.com)}
\thanks{F.J. de Bruijn and H.A.W. Schmeitz are with Philips Research, Eindhoven, The Netherlands. (e-mails: frits.de.bruijn@philips.com, harold.schmeitz@philips.com)}
\thanks{Y.C. Eldar is with the faculty of Math and Computer Science, Weizmann Institute of Science, Rehovot, Israel. (e-mail: yonina.eldar@weizmann.ac.il)}}

\maketitle

\begin{abstract}
Biomedical imaging is unequivocally dependent on the ability to reconstruct interpretable and high-quality images from acquired sensor data. This reconstruction process is pivotal across many applications, spanning from magnetic resonance imaging to ultrasound imaging. While advanced data-adaptive reconstruction methods can recover much higher image quality than traditional approaches, their implementation often poses a high computational burden. In ultrasound imaging, this burden is significant, especially when striving for low-cost systems, and has motivated the development of high-resolution and high-contrast adaptive beamforming methods.
Here we show that deep neural networks that adopt the algorithmic structure and constraints of adaptive signal processing techniques can efficiently learn to perform fast high-quality ultrasound beamforming using very little training data. We apply our technique to two distinct ultrasound acquisition strategies (plane wave, and synthetic aperture), and demonstrate that high image quality can be maintained when measuring at low data-rates, using undersampled array designs. Beyond biomedical imaging, we expect that the proposed deep~learning based adaptive processing framework can benefit a variety of array and signal processing applications, in particular when data-efficiency and robustness are of importance.

\end{abstract}

\flushbottom
\maketitle

\thispagestyle{empty}

\section{Introduction}

\noindent The reconstruction of high-quality images from measured sensor data is essential for diagnostic imaging, facilitating the timely diagnosis and treatment of life threatening diseases and aiding in the personalization of patient care. Unfortunately, current high-quality imaging technologies have major drawbacks: magnetic resonance imaging (MRI) is highly expensive, and X-ray computed tomography (CT) induces harmful ionizing radiation. 

Conversely, ultrasound imaging has been an invaluable diagnostic tool due to its low cost, portability and generally causing little discomfort to the patient by being minimally invasive and free of ionizing radiation. Additionally, its high degree of interactability is unique among the various imaging options, enabled by real-time imaging using efficient image reconstruction algorithms. Unfortunately this technique is also known for reduced image quality as compared to other imaging modalities.

\begin{figure*}[t]
	\centering
	\includegraphics[width=17cm, trim={0cm 0.0cm 0cm 0.0cm},clip]{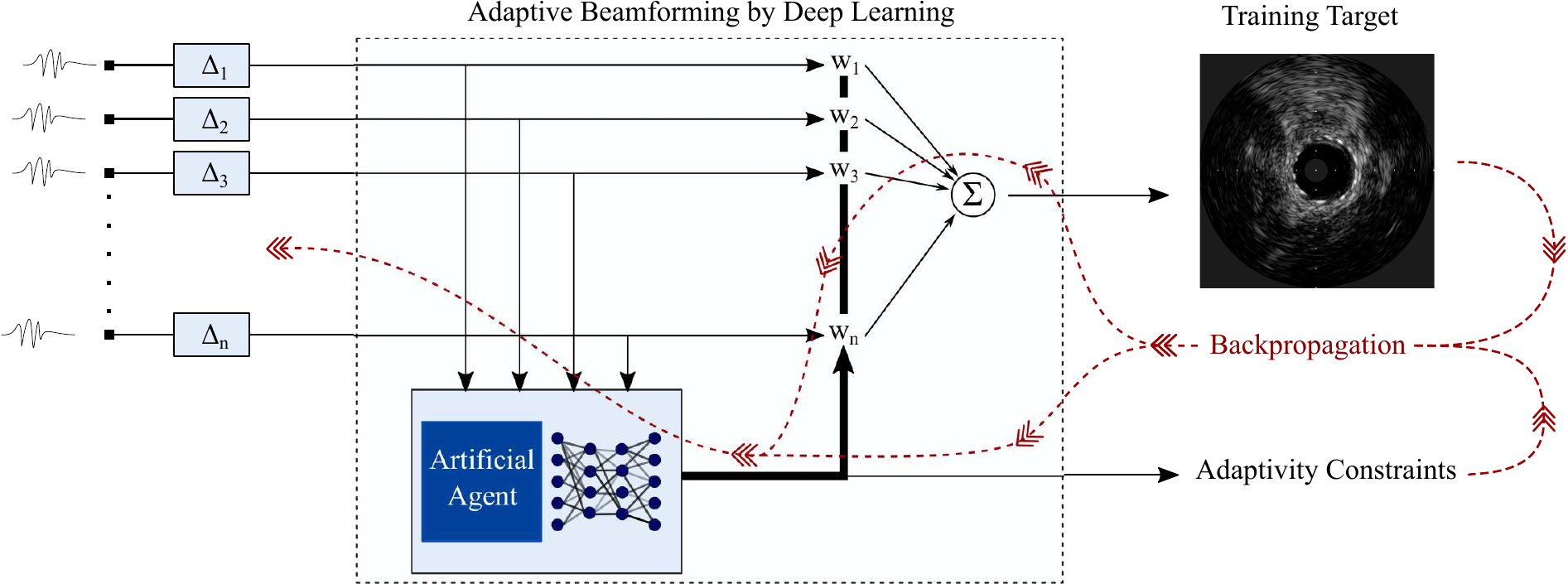}
	\caption{Schematic overview of \acro. Using the time aligned array signals as input, an artificial agent calculates an optimal set of apodization weights. The time-aligned array signals are subsequently multiplied by the predicted weights and summed to yield a beamformed output signal. This optimality is learned through training by backpropagating over a training target and adaptivity constraints. }
	\label{fig:overview}
\end{figure*}

In ultrasound imaging, focused sonification strategies have been the standard for many years, where narrow scan lines are sequentially fired by applying focusing time delays to the active transducer elements. The resulting backscattered echoes are then processed to reconstruct a sub-region of the image. The total scan time consequently increases linearly with the amount of acquired scan lines and the physical time-of-flight, which is governed by the speed of sound and desired imaging depth. This poses a trade-off between the achievable frame-rate and lateral image resolution. As a result, this limits lateral resolution in applications where high frame rates are favored. 

Recently, there has been a shift towards high-frame-rate unfocused sonification techniques such as Plane Wave (PW) and Synthetic Aperture (SA) imaging. 

Due to their unfocused transmission, these methods rely more heavily on digital reconstruction algorithms, where dynamic focusing is applied upon reception. In commercial devices, to maintain real-time reconstruction rates, this digital processing is done using a low-complexity, static delay-and-sum (DAS) beamformer, which applies predetermined delays and weights to the channel signals, after which the individual contributions are summed to yield a beamformed signal. This inherently poses a compromise between main lobe width and side lobe intensity, or equivalently, resolution and contrast. The latter is especially affected during unfocused transmissions, where clutter is received from a wide insonified area. 

Considerable improvements in image quality have been achieved through data-adaptive beamforming methods, of which Minimum Variance (MV) Distortionless Response \cite{Capon1969} \cite{Holfort2009} is one of the most common approaches. This technique achieves increased resolution without compromising contrast by adaptively placing sidelobes in directions where little energy is received. However, MV beamforming is computationally much more demanding than DAS, resulting in a significantly longer reconstruction time and thereby depriving it of the interactability that makes ultrasound imaging so appealing compared to e.g. MRI and CT. While significant progress has been made to decrease the computational time of MV beamforming algorithms \cite{Kim2014}\cite{Bae2016}, real-time implementation remains a major challenge.

Several lower complexity solutions have been proposed to improve image quality. Coherence Factor (CF) weighing \cite{Mallart1994} aims to reduce clutter by multiplying the DAS beamformed output with a coherence factor, defined as the ratio between the coherent and incoherent energy across the aperture. This method however suffers from artifacts when the SNR is low, due to the reduced accuracy of the coherent energy estimation \cite{Nilsen2010}. For broad insonification techniques such as PW and SA imaging, this is especially problematic. 

\noindent In \cite{Chernyakova2019}, Chernyakova \textit{et. al.} proposed an iterative maximum-a-posteriori (iMAP) estimator, which provides a statistical interpretation to beamforming by treating the signal of interest and interference as uncorrelated Gaussian random variables. The method has shown to achieve higher contrast as compared to other low-complexity beamformers described in the literature. A different approach based on the convolution of the delayed RF signals, named COBA, was presented in \cite{Cohen2018}. This technique has led to better lateral resolution and contrast, while only slightly increasing complexity, as it can be implemented efficiently using the fast Fourier transform.

Meanwhile, deep learning has emerged as a popular tool to solve inverse problems, leading to revolutionary breakthroughs in domains ranging from computer vision to natural language processing. Deep neural networks (DNNs) are trained to perform advanced tasks based on large amounts of data. DNNs consist of many layers of interconnected artificial neurons, which on their own perform only simple operations, but when combined are universal function approximators~\cite{Hornik1989}. Once trained, inference is typically fast, especially on GPU accelerated systems. DNNs have proven to be very successful in tasks such as image classification~\cite{Krizhevsky2012}, image segmentation~\cite{Garcia2017} and speech recognition~\cite{Hinton2012}.

Naturally, these techniques are also receiving significant attention in medical imaging. Although the focus has mainly been on solving image analysis tasks such as classification and segmentation~\cite{Greenspan2016}, more recent developments exploit deep learning for the image reconstruction process itself, finding application in X-ray CT, MRI, PET and photoacoustic tomography~\cite{Wang2018}. An overview of the application of deep learning in various ultrasound imaging tasks can be found in~\cite{vanSloun2019DL_in_US}. DNNs have been used for interpolating missing radio frequency (RF) data~\cite{Yoon2017}, reducing off-axis scattering~\cite{Luchies2018}, compounding of plane waves~\cite{Gasse2017} and super-resolution microscopy~\cite{vanSloun2018}. In~\cite{Nair2018}, the beamforming step was circumvented altogether, achieving direct segmentation of cyst phantoms from simulated RF channel data. For many clinical applications however, a brightness-mode (B-mode) image is desired, though at a higher resolution and contrast than that achieved with conventional DAS beamforming. 

Deep learning strategies are strongly data dependent, relying on vast and diverse training datasets accompanied by a sufficiently large network in order to find a correct and generalized mapping from input to output. Even after convergence, it is hard to prove the robustness of such a network for new, unforeseen data, a trait that is undesirable in a clinical setting where a predictable system response is of vital importance for correct diagnosis of diseases. 

Rather than relying on a large general-purpose network in conjuncture with abundant training data, we aim to train a network that is specifically designed for a certain task, thereby limiting its degrees of freedom, and allowing for a more compact architecture that is data efficient in training. We here propose Adaptive Beamforming by deep LEarning (\acro), a method that exploits the algorithmic structure of adaptive beamforming, with a neural network adaptively computing a set of optimal image reconstruction parameters given the received RF data (see Fig. \ref{fig:overview}) \cite{luijten2019deep}. In this case, these parameters constitute the adaptive apodization weights, as in MV beamforming. 

We demonstrate the performance of \acro{} on two widely different ultrasound imaging modalities: PW (linear array, external) and SA (circular array, intravascular) imaging. Intravascular ultrasound imaging (IVUS) is an invaluable tool for e.g. interventional cardiography. Due to the strong constraints on size and power consumption of the cathether-based probe, this method is particularly susceptible to noise and grating lobe artifacts, making it an excellent candidate to test the merit of our adaptive deep-learning-based beamformer. We then show that \acro{} can also learn to cope with subsampled SA transmit designs. The image quality of ABLE is compared against three existing beamformers: DAS, iMAP and MV. Furthermore we assess the computational complexity of these methods to highlight its implication on real-time performance.

While the scope of this work concerns ultrasound beamforming, we emphasize that this method is generally applicable in other applications requiring content-adaptive prediction of signal processing parameters. 

The remainder of this paper is organised as follows. In Section~II we describe existing beamforming methods. In Section~III we introduce ABLE and elaborate on its design and training strategy. Data acquisition, validation of image quality, complexity analysis and subsampling are discussed in Section~IV. In Section~V, we present experimental results, which are discussed in Section~VI. We conclude in Section~VII. 

\section{Existing Beamformers}

\subsection{Delay and Sum Beamforming}

\noindent In ultrasound beamforming, the received channel signals are first time-of-flight corrected by applying delays according to the geometry of the transducer and imaging scene. Pixel-wise time delays for each array channel can be calculated as:
\begin{equation}
    \Delta[x,z] = \Delta[\mathbf{r}] = \frac{||\mathbf{r}_{TX}-\mathbf{r}||_2+||\mathbf{r}_{RX}-\mathbf{r}||_2}{c},
\end{equation}
where $\Delta[\mathbf{r}]$ is the required channel delay to focus to an imaging point $\mathbf{r}$, $\mathbf{r}_{TX}$ and $\mathbf{r}_{RX}$ are the positions of the transmitting and receiving element in the array, respectively, and $c$ is the speed of sound in the medium. 
By dynamically focusing to different imaging points, corresponding to pixels in the final image, a pixel-wise response can be generated for every receiving array element $n\in~\{1, ..., N\}$, given by
\begin{equation}
\mathbf{y}_n[x,z] = \mathbf{x}_n[\Delta([x,z])],
\end{equation}
where $\mathbf{x}_n[t]$ denotes the channel signal for the $n^\text{th}$ element.

A DAS-beamformed image $\mathbf{P}_{\text{DAS}}[x,z]$ is constructed by multiplying the focused contributions of each receiving channel with predetermined (but possibly spatially varying) weights $\mathbf{w}[x,z]$, and summing the result:
\begin{equation}
\mathbf{P_{\text{DAS}}}[x,z] = \mathbf{w}^\textrm{H}[x,z] \mathbf{y}[x,z].
\end{equation}
Often, these weights are set according to a boxcar ($\mathbf{w} = \mathbf{1}^N$), Hanning or tapered cosine window.

\subsection{Iterative Maximum-A-Posteriori}

The iMAP algorithm \cite{Chernyakova2019} provides a reweighing of the DAS beamformed output for $\mathbf{w} = \mathbf{1}^N$, calculated in two iterative steps. First, the signal and noise variances are estimated by
\begin{equation}
    \{\hat{\sigma^2_x}, \hat{\sigma^2_n\}}_{(t)} = \left\{\mathbf{\hat{P}}^2_{\text{iMAP},(t)}, \frac{1}{N}||\mathbf{y}-\mathbf{\hat{P}}_{\text{iMAP},(t)}\mathbf{1}||^2\right\},
    \label{eqn:imap_rule1}
\end{equation}
and initializing $\mathbf{\hat{P}}_{\text{iMAP},(0)} = \mathbf{1}^\textrm{H} \mathbf{y}$, where $(\cdot)^\textrm{H}$ stands for the Hermitian transpose. 
Using these variances, a MAP estimate of the beamformed signal is given by
\begin{equation}
    \mathbf{\hat{P}}_{\text{iMAP},(t+1)} = \frac{\hat{\sigma}^2_{x,(t)}}{\hat{\sigma}^2_{x,(t)} + N\hat{\sigma}^2_{x,(t)}} \mathbf{1}^\textrm{H}\mathbf{y}.
    \label{eqn:imap_rule2}
\end{equation}
Equation (\ref{eqn:imap_rule1}) and (\ref{eqn:imap_rule2}) are iterated until a stopping criterion is met. In the rest of the paper we will consider iMAP2, meaning the algorithm using 2 interations.

\subsection{Adaptive Eigen-Based Minimum Variance Beamforming}
\noindent The MV beamformer provides a set of content-adaptive apodization weights for each pixel; these weights are selected to minimize output power (variance) of the beamformed signal, subject to a distortionless response in the desired direction.
Accordingly, the following minimization problem is solved for every pixel in the image:
\begin{equation}
\begin{aligned}
\min_\mathbf{w}\mathbf{w}^\textrm{H}[x,z]\mathbf{R}[x,z]\mathbf{w}[x,z]\\
s.t.\quad \mathbf{w}^\textrm{H}[x,z]\mathbf{a} = 1
\end{aligned}
\label{eqn:mv}
\end{equation}
where $\textbf{R}$ denotes the sample covariance matrix calculated across the delayed receiving array signals, and $\mathbf{a}$ is a steering vector. Since the data is already time-of-flight corrected, we select $\mathbf{a} = \mathbf{1}^N$. Solving (\ref{eqn:mv}) yields the following closed form solution:
\begin{equation}
\mathbf{w}_{\text{MV}}[x,z] = \frac{\mathbf{R}^{-1}[x,z]\mathbf{a}}{\mathbf{a}^\text{H}\mathbf{R}^{-1}[x,z]\mathbf{a}}.
\label{eqn:mvclosedform}
\end{equation}
With the aim of avoiding a potentially unstable numerical inversion of $\mathbf{R}$, the covariance matrix is estimated by applying spatial smoothing using a subaperture $\tilde{\mathbf{y}}_l[x,z] \in \mathbb{R}^L$  of the full response $\mathbf{y}[x,z]$:
\begin{equation}
\mathbf{\hat{R}}[x,z] = \frac{ \sum_{l = 0}^{N-L}\tilde{\mathbf{y}}_l[x,z]\tilde{\mathbf{y}}_l^\textrm{H}[x,z]}{N-L+1}.
\end{equation}
where
\begin{equation}
\mathbf{y}_l[x,z] = \Big[ y_l[x,z], \quad y_{l+1}[x,z], \quad  ...  \quad y_{l+L-1}[x,z] \Big]
\end{equation}
To further improve stability, diagonal loading is applied by adding a factor $\epsilon$ to the diagonal of $\mathbf{\hat{R}}$ determined by the trace operator as:
\begin{equation}
    \epsilon = D\cdot trace(\mathbf{\hat{R}}),
\end{equation}
where $D$ is the diagonal loading factor. The resulting solution for $\mathbf{w}$ corresponds to regularized MV and is computationally dominated by the inversion of the $N\times N$ covariance matrix~$\mathbf{R}$. 

Additional improvements in image contrast can be achieved by employing an Eigen-Based Minimum Variance Beamformer (EBMV) \cite{Deylami2016}, at the cost of higher computational complexity. Assuming that the received signal is a combination of desired signal components and added noise, we can take the eigen-decomposition of $\mathbf{R}$ and project the signal subspace $\mathbf{E}_\text{signal}$, composed of the dominant, desired eigenvectors, on the weight vector to obtain the updated weights:
\begin{equation}
\label{eqn:eigproj}
\mathbf{w_{\text{EBMV}}} = \mathbf{E_{\text{signal}}}\mathbf{E}^\textrm{H}_{\text{signal}} \mathbf{w}.
\end{equation}
This procedure is repeated for each pixel to obtain a EBMV-beamformed image:
\begin{equation}
\mathbf{P_{\text{EBMV}}}[x,z] = \mathbf{w}_\text{EBMV}^\textrm{H}[x,z] \mathbf{y}[x,z].
\end{equation}

\section{ABLE: Adaptive Beamforming by Deep Learning}

\subsection{Network Architecture}
\noindent \acro's architecture is inspired by the MV beamformer, with a neural network that adaptively calculates apodization weights according to the input channel data. Thus, the traditional computationally expensive adaptive processor is replaced with a more computationally efficient and more powerful artificial agent in the form of a neural network, as shown in Fig. \ref{fig:overview}. This specific design allows the artificial agent to adaptively tune a set of image reconstruction parameters, in this case the apodization weights, within a predictable beamforming structure. A beamformed pixel is obtained by multiplying the time-of-flight corrected array channel data for that pixel with the apodization weights as produced by the neural network, $f_\theta:\mathbb{R}^N\rightarrow\mathbb{R}^N$, such that:
\begin{equation}
\mathbf{P_{\text{ABLE}}}[x,z] = f_\theta(\mathbf{y}[x,z])^\textrm{H} \mathbf{y}[x,z].
\end{equation}

\begin{figure}[t]
	\centering
	\includegraphics[width=8.7cm]{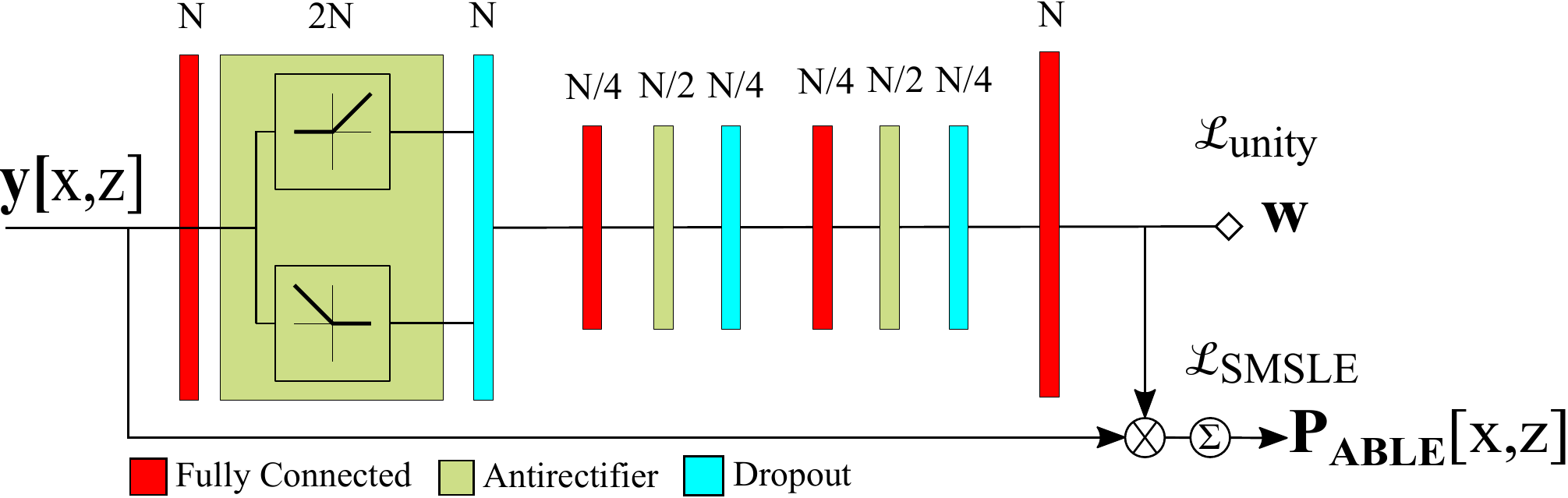}
	\caption{Schematic overview of the proposed neural network. Above each layer the number of output nodes is denoted as a function of the aperture size $N$.}
	\label{fig:network}
\end{figure}

The proposed network $f_\theta$ consists of 4 fully connected (FC) layers comprising $N$ nodes for the outer layers, corresponding to the receiving aperture size, and $N/4$ nodes for the inner layers, as indicated in Fig. \ref{fig:network}. The latter introduces a compact latent space which forces the network to find a compact representation of the data, aiding in noise suppression. Between every FC layer, dropout is applied, randomly dropping nodes with a probability of 0.2, to prevent overfitting.

\begin{figure*}
\centering
\begin{minipage}{.49\textwidth}
  \centering
  \includegraphics[width=1.00\linewidth]{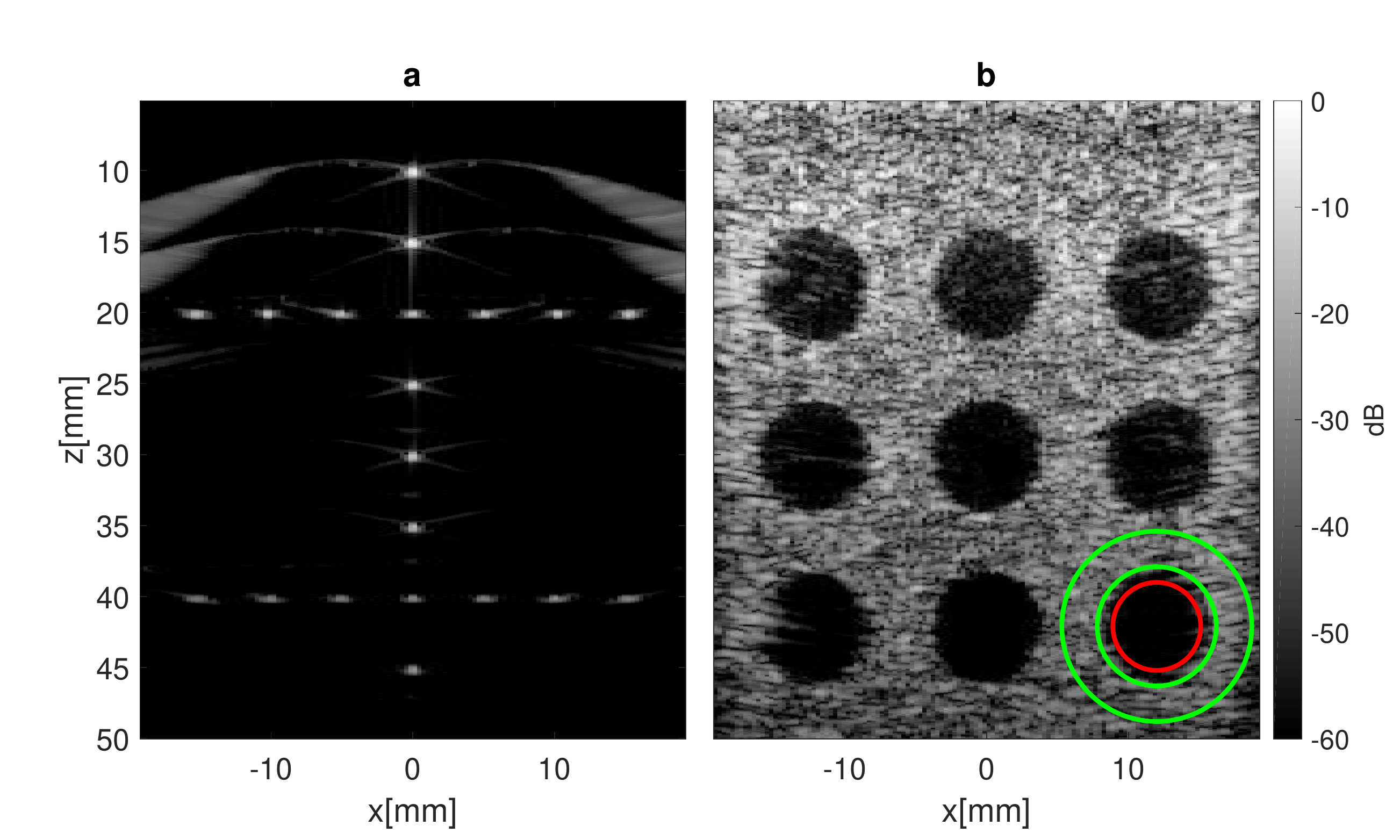}
\end{minipage}%
\begin{minipage}{.49\textwidth}
  \centering
  \includegraphics[width=1.00\linewidth]{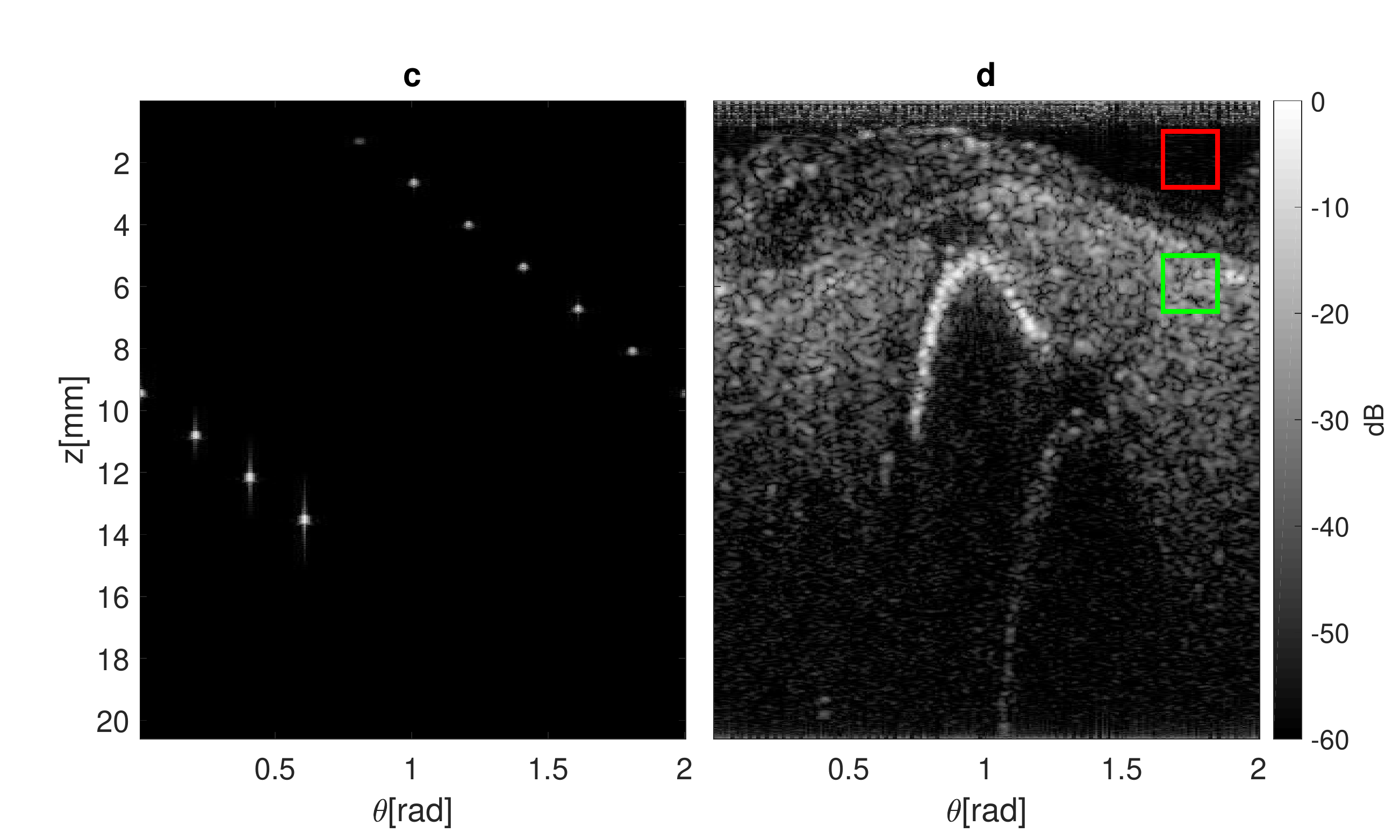}
\end{minipage}
\caption{Images used for the performance evaluation of the beamforming methods. \acro{}'s reconstructions are shown for: a) Simulated point scatterers (plane wave, linear array) b) Simulated anechoic cysts (plane wave, linear array) c) Simulated point scatterers (synthetic aperture, IVUS circular array) d) \textit{In-vivo} coronary artery (synthetic aperture, IVUS circular array). For contrast evaluation, high and low intensity regions are indicated in green and red, respectively.}
\label{fig:evaluation}
\end{figure*}

\subsection{Activation functions}

\noindent Rectified linear units (ReLUs) are the most commonly used activations in DNNs because of their computational efficiency, ability to provide sparse representations and largely avoiding vanishing gradients due to a positive unbounded output \cite{LeCun2015}.
However, when dealing with RF data, containing positive and negative values, such a non-linearity may not be appropriate, as it inherently leads to many `dying' nodes, impairing the training process. In contrast, a hyperbolic tangent activation is able to preserve negative values. It is however bounded between -1 and 1, and therefore tends to saturate quickly for signals with a large dynamic range, resulting in a vanishing gradient during back propagation. This behavior becomes especially problematic in DNNs because of the substantial amount of consecutive activations \cite{Glorot2010}.

Instead, we propose to use an antirectifier layer\footnote{Fran{\c{c}}ois Chollet, Antirectifier, Github, https://github.com/keras-team/keras/blob/master/examples/antirectifier.py}, which combines a sample wise $\ell_2$ normalization with the concatenation of the positive and the negative part of the input. This operation is described as:
\begin{equation}
g(\mathbf{x}) = 
\begin{bmatrix}
\max\left(0,\frac{\mathbf{x}-\mu_x}{\lVert \mathbf{x} - \mu_x \rVert_2}\right)\\
\max\left(0,-\frac{\mathbf{x}-\mu_x}{\lVert \mathbf{x} - \mu_x \rVert_2}\right) 
\end{bmatrix},
\end{equation}
where $\mu_x$ denotes the mean of $\mathbf{x}$. The antirectifier effectively introduces non-linearity, while preserving the contributions of negative signal components as well as the dynamic range of the input.

\subsection{Training Strategy}
\noindent Our network is implemented in Python using the Keras API with a Tensorflow (Google, CA, USA) backend on a compute cluster with an Nvidia Tesla K80 graphics card. For training, the Adam optimizer was used with a learning rate of 0.001, stochastically optimizing the loss using batches that each contain all pixels belonging to a single image. A randomized training strategy was employed, where the data sequence supplied to the network was randomized at each epoch. The total training loss comprised an image loss $\mathcal{L}_{\text{SMSLE}}$, and an apodization-weight penalty $\mathcal{L}_{\text{unity}}$, leading to:
\begin{equation}
\mathcal{L}_{\text{total}} = \lambda\mathcal{L}_{\text{SMSLE}} + (1-\lambda) \mathcal{L}_{\text{unity}},
\end{equation}
where $\lambda$ is a parameter between 0 and 1 which determines the weight ratio of the losses. We detail on the individual loss functions in the following.

\subsubsection{Image Loss}
The image loss is designed to promote similarity between the target image and the image produced using \acro. Given the large dynamic range of ultrasound backscattering, ultrasound images are typically logarithmically compressed to create a visually more appealing and insightful image. Consequently, conventional loss metrics (e.g. mean-square-error (MSE) or mean-absolute-error (MAE)) do not provide a balanced representation of deviation from desireable image properties. 
To better reflect this, we introduce the signed-mean-squared-logarithmic error (SMSLE) as a loss function. The beamformed data is split into a positive ($\mathbf{P}^+$) and negative ($\mathbf{P}^-$) part on which the mean-squared-logarithmic-error is calculated. The total loss is the sum of these two contributions, resulting in:
\begin{multline}
\mathcal{L}_{\text{SMSLE}} = \frac{1}{2}\lVert\log_{10}(\mathbf{P}_\text{ABLE}^+)-\log_{10}(\mathbf{P}_\text{MV}^+)\rVert_2^2\quad+ \\
					\frac{1}{2}\lVert\log_{10}(-\mathbf{P}_\text{ABLE}^-)-\log_{10}(-\mathbf{P}_\text{MV}^-)\rVert_2^2.
\end{multline}

\subsubsection{Weight Penalty}
The direct access to the generated apodization weights permits additional penalties to be invoked during training. Inspired by MV beamforming, we promote a distortionless response by penalizing deviations from unity gain using:
\begin{equation}
\mathcal{L}_{\text{unity}} = |\mathbf{1}^T f_\theta(\mathbf{y}[x,z]) - 1|^2.
\end{equation}

\section{Experiment Setup}

\subsection{Data Acquisition}
\noindent We acquired both PW (linear array) and SA (IVUS) datasets, each comprising 1000 imaging frames. In addition to this, 100 simulated frames of point scatterers were generated for each acquisition method. 
\textit{In-vivo} PW images of the carotid artery of a single person were acquired using the Vantage system (Verasonics Inc., WA, USA) in combination with the \mbox{L11-4v} linear transducer. SA IVUS images of an \textit{in-vivo} coronary artery of a single patient, an \textit{in-vitro} wire-target phantom and arterial stent, were recorded using the EEP PV014P circular transducer (Philips Volcano, USA). In this particular system, each transmit is received by an aperture of 14 elements. During beamforming, a virtual aperture is constructed by combining the received signals of multiple transmits and receives. 
The transducer parameters for the respective acquisition methods are indicated in Table~\ref{tab:params}. 
For validation purposes, a separate hold-out test set was taken from the full dataset, supplemented with simulated PW images from the PICMUS dataset \cite{Liebgott2016}.
\begin{table}
	\renewcommand{\arraystretch}{1.2}
	\centering
	\caption{Transducer Parameters}
	\label{table:par}
	\begin{tabular}{c|c c}
		\textbf{Parameter}&\textbf{Plane wave}&\textbf{Synthetic aperture}\\
		\hline
		\hline
		Elements & 128 & 64\\
		Aperture size & 128 & 105 (virtual)\\
		Pitch & 0.300 mm & 0.057 mm\\
		Transmit Frequency & 6.25 MHz & 20MHz\\
		Sampling Frequency & 25 MHz & 100MHz\\
	\end{tabular}
	\label{tab:params}
\end{table}

The acquired data is used to train \acro{} to map time-of-flight corrected RF input signals to corresponding high-quality beamformed targets. For the latter we employ an empirically fine-tuned EBMV beamformer. 

\subsection{Validation}
Resolution was assessed by evaluating the Full-Width-at-Half-Maxima (FWHM) of simulated point scatterers, which were calculated for both the lateral and the axial direction. For the circular-array based SA scans, lateral resolution was evaluated in the polar domain. An example of these simulated point-scatterers can be seen in Fig.~\ref{fig:evaluation}a and \ref{fig:evaluation}c. 

Contrast of the images was measured by the contrast-to-noise ratio (CNR) between a low and high intensity region, defined as:
\begin{equation}
\text{CNR} = 20\log_{10}\left(\frac{|\mu_{low}-\mu_{high}|}{\sqrt{(\sigma_{low}^2+\sigma_{high}^2)/2}}\right),
\end{equation}
where $\mu_{low}$ and $\mu_{high}$ represent the mean intensity of the low and high intensity regions, respectively, and $\sigma_{low}^2$ and $\sigma_{high}^2$ the variance of the low and high intensity regions, respectively. The CNR was evaluated on simulated anechoic cysts for PW imaging with a linear array, and an \textit{in-vivo} coronary artery for SA IVUS imaging (see Fig.~\ref{fig:evaluation}b and \ref{fig:evaluation}d).

\begin{figure*}[]
	\centering
	\includegraphics[width=16.8cm, trim={2.9cm 0.1cm 1.3cm 0cm},clip]{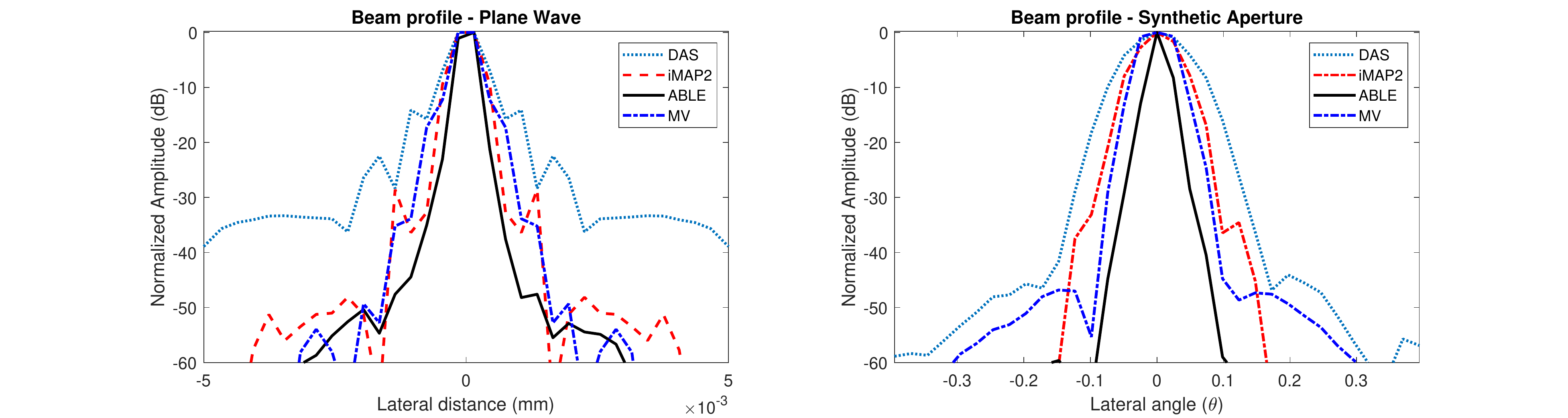}
	\caption{Normalized beamprofiles along a simulated pointscatterer for (left) Plane-wave imaging and (right) Synthetic Aperture (IVUS) imaging.}
	\label{fig:beampattern}
\end{figure*}

\subsection{Algorithmic Complexity}

Here we assess the computational complexity of the adaptive beamformers described here, in terms of floating-point-operations (FLOPs). Since all beamformers require time-of-flight correction, and multiplication and summation of the array signals, as in DAS, these terms are disregarded. In other words, only the adaptive complexity is taken into account. 

\subsubsection{iMAP}

Following equation (\ref{eqn:imap_rule1}) and (\ref{eqn:imap_rule2}), we observe that for every iMAP iterations an estimate of signal and noise variances, and an update of the MAP estimate is required. Correspondingly, the total complexity amounts to 
\begin{equation}
    C_{\text{iMAP}} = I(\underbrace{3N+2}_{\text{variance estimation}} + \underbrace{3}_{\text{MAP estimate}})= I(3N+5)
    \label{eqn:CiMAP}
\end{equation}
where $I$ denotes the number of iterations.

\subsubsection{MV Beamforming}
The complexity of regularized MV beamforming is largely dependent on the inversion of the covariance matrix $\mathbf{R}$, with a complexity of $\mathcal{O}(N^3)$ \cite{Lorenz2005}. Additionally, the calculation of the weight vector $\mathbf{w}$ as in (\ref{eqn:mvclosedform}) requires a vector-vector and a vector-scalar product, and two vector-matrix products. Consequently, the total number of FLOPs needed can be determined by:
\begin{equation}
    C_{\text{MV}} = \underbrace{N^3}_{\text{matrix inversion}} + \underbrace{2N^2+3N}_{\text{weight calculation}}.
    \label{eqn:Cmv}
\end{equation}

The EBMV beamformer requires an additional eigen-decomposition of complexity $N^3$, and a projection of the dominant eigenvectors as described in (\ref{eqn:eigproj}). This results in a total complexity of: 
\begin{equation}
    C_{\text{EBMV}} = C_{\text{MV}} + \underbrace{N^3}_{\text{eigen decomposition}} + \underbrace{kN^2+N^2}_{\text{subspace projection}}.
    \label{eqn:Cebmv}
\end{equation}
In our implementation we project approximately 50\% of the eigenvectors, i.e. $k=0.5$.

\subsubsection{ABLE}
We consider the complexity of \acro{} during inference (i.e. a forward pass). Every FC layer can be described as a matrix-vector product with an added bias $\mathbf{Ax}+\mathbf{b}$, where $\mathbf{A}$ is the incidence matrix containing the weights of the layer, $\mathbf{b}$ is the bias vector, and $\mathbf{x}$ the input vector. Given that $\mathbf{A}$ is an $N_i \times N_{i+1}$ matrix, where $N_i$ and $N_{i+1}$ respectively denote the amount of input and output nodes of the layer, and similarly $\mathbf{b}$ has $N_{i+1}$ elements, we can determine that a total of  $2N_i N_{i+1}+N_{i+1}$ FLOPs are required for a single FC layer \cite{Boyd2004}.

ReLU activation, involving a comparison and multiplication, requires 2 FLOPs.
As a result, for the antirectifier, which concatenates two ReLU activations, 4 FLOPs are needed per node. Since dropout is not active during inference, it does not add to the complexity of the deployed network. 
In total, for our $L$-layer network, the number of required FLOPs amounts to:
\begin{equation}
\label{eqn:complexity}
    C_{\acro} = \underbrace{2N_0N_1+N_{1}}_{\text{input}} + \sum^{L-1}_{i = 1}\underbrace{4N_iN_{i+1}+N_{i+1}}_{\text{FC layers}} + \underbrace{4N_{i}}_{\text{activation}}.
\end{equation}

\begin{figure}[t!]
	\centering
	\includegraphics[width=8.5cm, trim={4.2cm 9.0cm 10.5cm 1.1cm},clip]{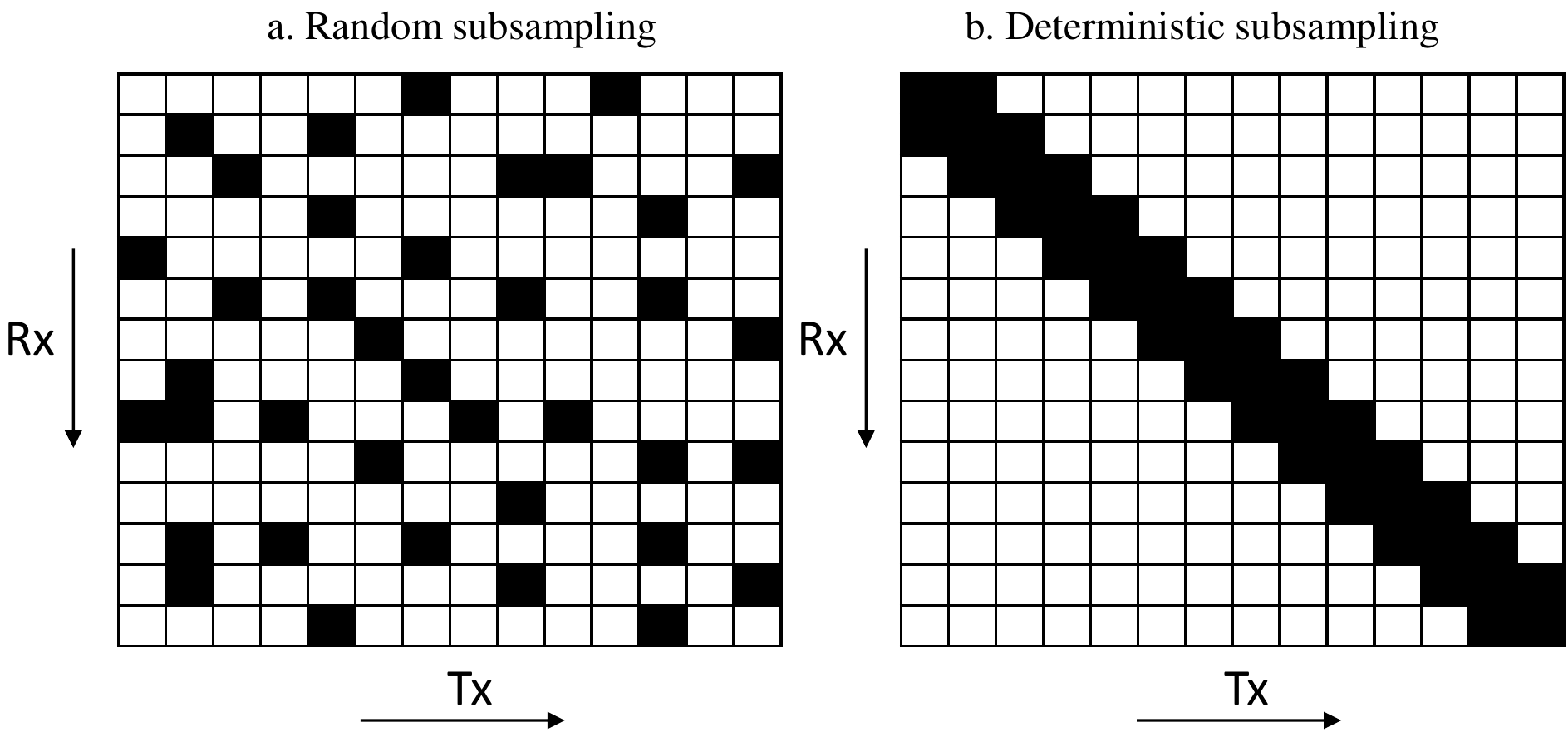}
	\caption{Visualization of the proposed subsampling schemes. Black boxes indicate active elements. a) Random subsampling, b) Deterministic subsampling.}
	\label{fig:subscheme}
\end{figure}

\begin{figure}[]
	\centering
	\includegraphics[width=7.6cm, trim={0.3cm 0cm 0.5cm 0.4cm},clip]{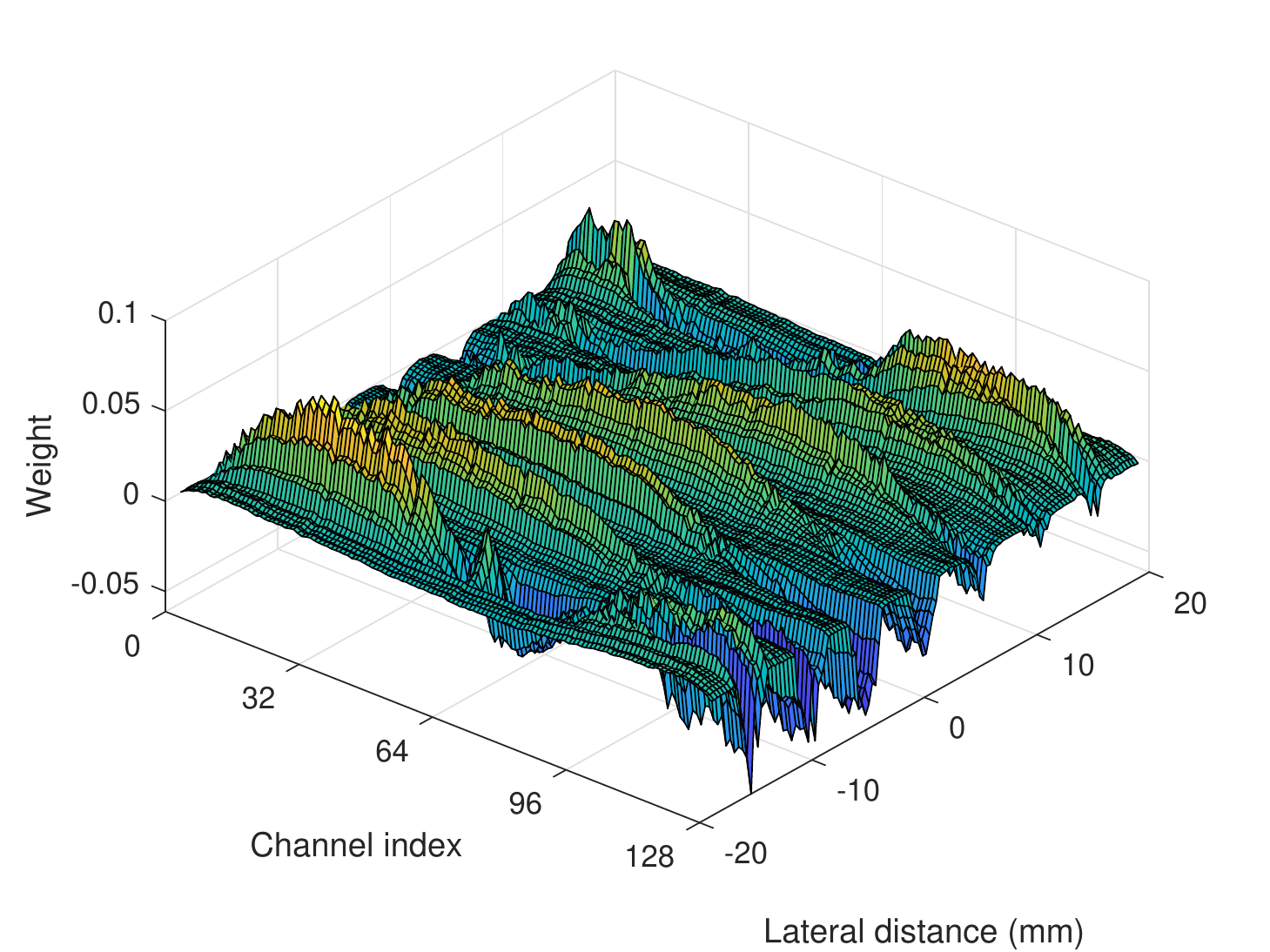}
	\caption{Visualization of the set of predicted apodization weights along a lateral line of 7 spaced point-scatterers.}
	\label{fig:weights}
\end{figure}

\subsection{Subsampling}

Finally we evaluate \acro's ability to reconstruct images from a subset of transducer elements in SA IVUS, using two different schemes (illustrated in Fig. \ref{fig:subscheme}). In the first, a subset of array elements is selected randomly from the full receiving array. In the second, the receiving elements that are close to the transmitting element are chosen, thereby favouring specular reflections.
We train the network to reconstruct images with subsampled array signals, at rates of 25\% and 50\%, as input, and the fully sampled EBMV reconstructions as target. 

\begin{figure*}[]
	\centering
	\includegraphics[width=17cm, trim={1cm 0.8cm 2cm 1.4cm},clip]{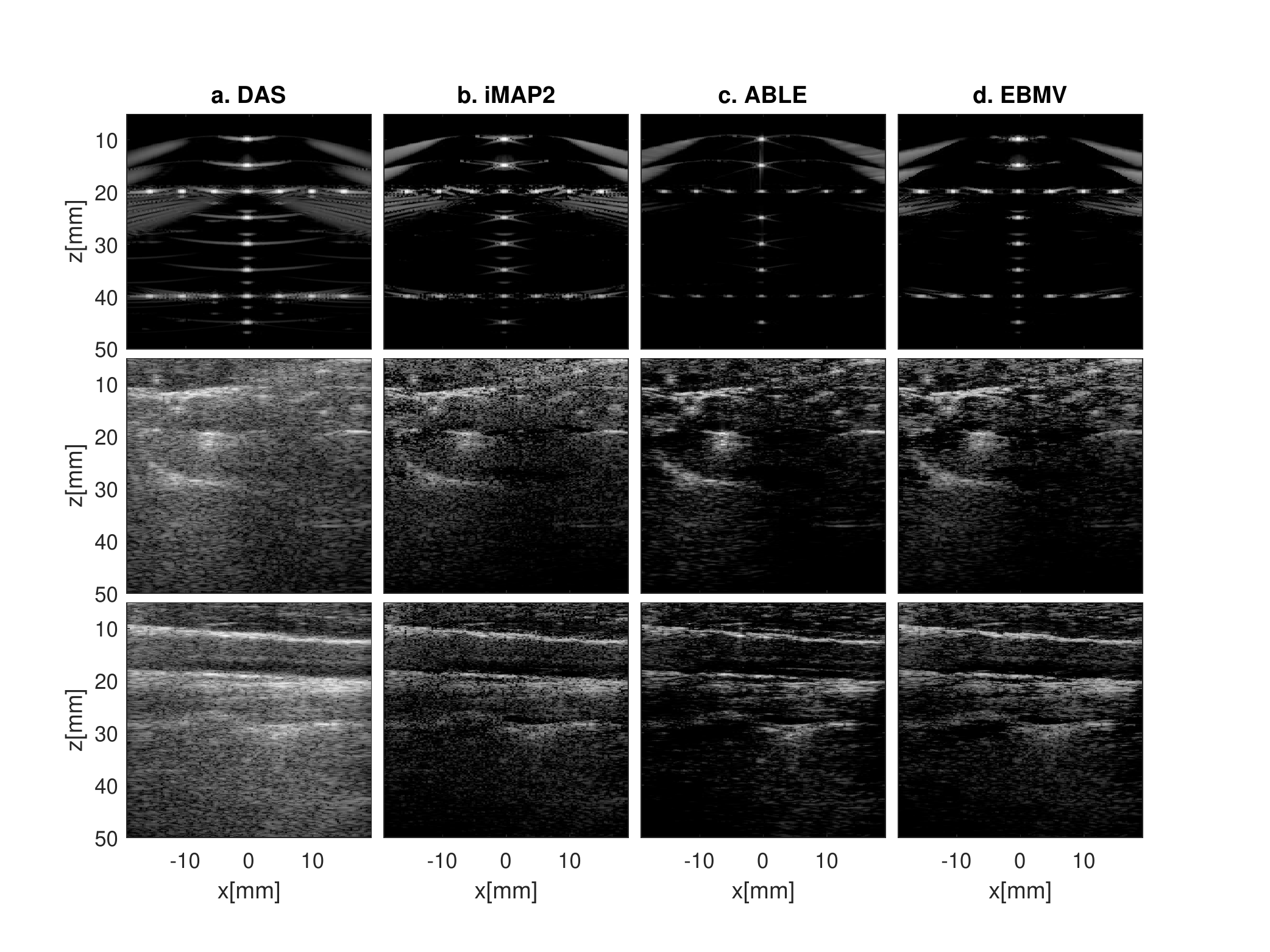}
	\caption{Single-plane-wave \textit{in-vivo} images using: a) Delay-and-sum (DAS) beamforming with Hanning apodization, b) iMAP2 beamforming c) Adaptive beamforming by deep learning (ABLE), and d) Eigen-Based Minimum variance (EBMV) beamforming. Images are logarithmically compressed with a dynamic range of 60dB. From top to bottom we have simulated point scatterers, carotid artery cross-section, and a carotid artery longitudinal cross-section}
	\label{fig:results_invivo}
\end{figure*}

\section{Results}

\noindent The obtained axial resolution, lateral resolution and CNR for the beamforming methods described in Section~II are given in Table~\ref{table:results}, showing that \acro{} outperforms conventional DAS and iMAP2 beamforming for both PW (linear array) and SA (IVUS) imaging. Their beam profiles are visualized in Fig.~\ref{fig:beampattern}. Interestingly, \acro{} even yields better lateral resolution than its train target, EBMV, displaying the power of mean-error optimization across many examples. 

Figure~\ref{fig:weights} exemplifies the apodization weights calculated by \acro{} along a lateral line containing a set of equally spaced point-scatterers (see Fig. \ref{fig:evaluation}a). The interpretable adaptivity of the apodization, and its relation with the spatial location of the point-scatterers is noteworthy.

Figure~\ref{fig:results_invivo} shows the PW images obtained using DAS, iMAP2, \acro{} and EBMV beamforming. From the simulated point scatters one can see that DAS beamforming yields wider and more pronounced sidelobes than iMAP2, \acro{} and EBMV. The \textit{in-vivo} images show strong clutter suppression for \acro{} and EBMV compared to DAS.

\begin{table}[b!]
\small
	\renewcommand{\arraystretch}{1.2}
	\centering
	\caption{Resolution and contrast metrics}
	\label{table:results}	
	\begin{tabular}{c c c c c}
	    \multicolumn{4}{c}{\textbf{Plane wave (linear array)}}\\
		& DAS & iMAP2 & \acro & MV\\
		\hline
		\hline

		$\text{FWHM}_{\text{lat}}$ (mm) & 0.846 & 0.730 & 0.704 & 0.778\\
		$\text{FWHM}_{\text{ax}}$ (mm) & 0.431 & 0.430 & 0.342 & 0.434\\
		$\text{CNR}$ (dB) & 11.35 &  9.59 & 11.91 & 12.67\\

		\multicolumn{4}{c}{\textbf{}}\\
		\multicolumn{4}{c}{\textbf{Synthetic Aperture (circular array)}}\\
		& DAS & iMAP2 & \acro & MV\\
		\hline
		\hline
		$\text{FWHM}_{\text{lat}}$ (rad) & 0.127 & 0.094 & 0.055 & 0.087\\
		$\text{FWHM}_{\text{ax}}$ (mm) & 0.217 & 0.173 & 0.118 &  0.046\\
		$\text{CNR}$ (dB) & 9.96 & 8.95 & 13.44 & 13.03\\
	\end{tabular}
\end{table}

\begin{figure*}[]
	\centering
	\includegraphics[width=17cm, trim={1cm 0.8cm 1cm 1.4cm},clip]{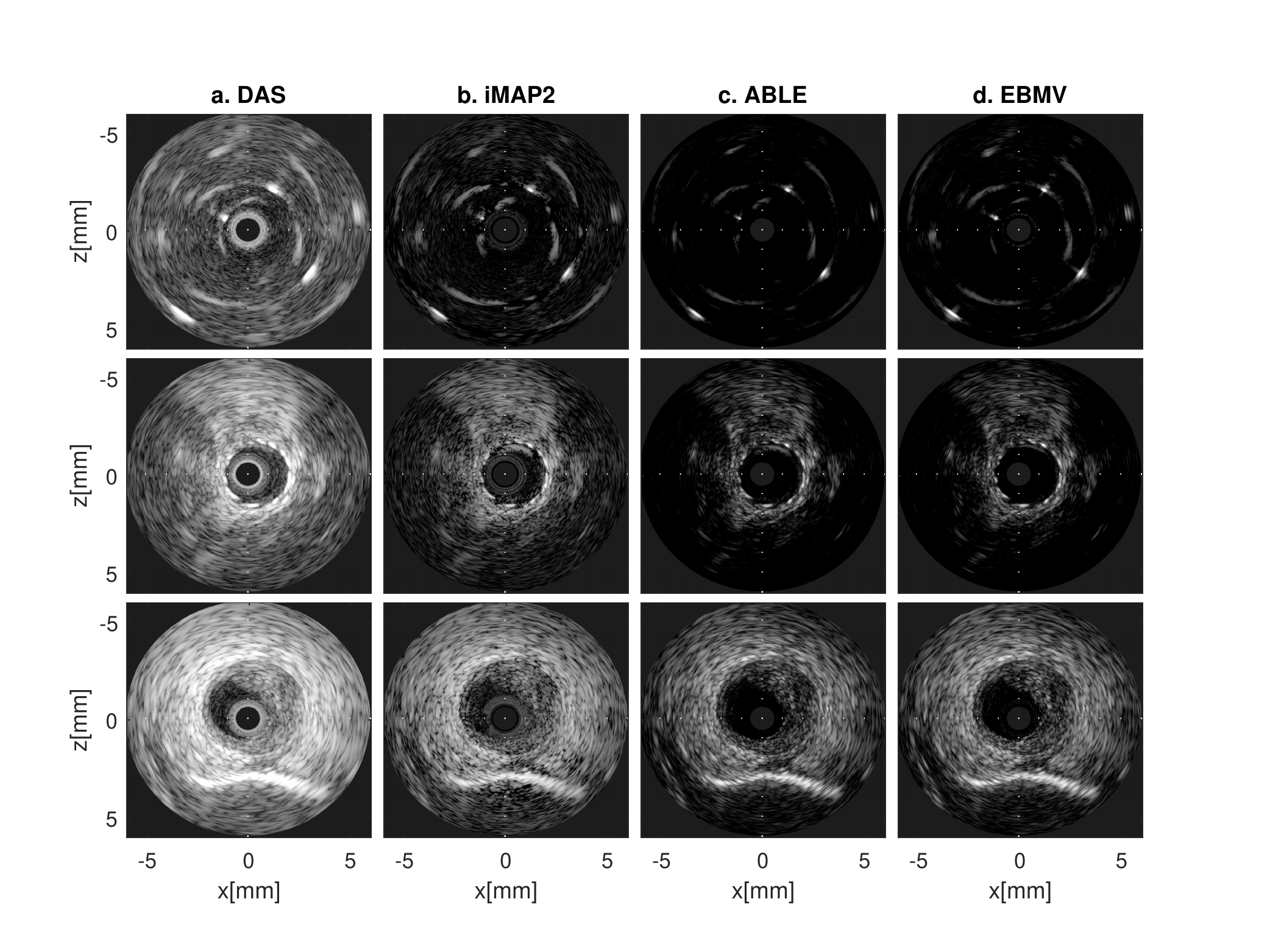}
	\caption{Reconstructed Synthetic Aperture \textit{in-vivo} images using: a) Delay-and-sum (DAS) beamforming with Hanning apodization, b) iMAP2 beamforming c) Adaptive beamforming by deep learning (ABLE), and d) Eigen-Based Minimum variance (EBMV) beamforming. Images are logarithmically compressed with a dynamic range of 60dB. From top to bottom the we have a wire phantom, an arterial stent, and a coronary artery with plaque.}
	\label{fig:results_ivus_invivo}
\end{figure*}

The reconstructed \textit{in-vivo} SA IVUS images are given in Fig.~\ref{fig:results_ivus_invivo}. We again observe a notable reduction in sidelobe intensity near the simulated point scatterers for \acro{} and EBMV, as compared to DAS. Consequently, in the \textit{ex-vivo} arterial stent image, we see a more pronounced distinction between tissue and the stent struts. Substructures in the coronary plaque also become more clearly visible due to reduced clutter and improved resolution. 

Next, we asses the computational complexity of the beamformers. From~(\ref{eqn:complexity}), we determine that \acro{} requires 71,232 and 47,930 FLOPs to calculate the apodization weights for the linear array ($N=128$) and the IVUS circular array ($N=105$), respectively. This is well below the 6,186,599 and 414,994 FLOPs required for EBMV beamforming on those systems. In practice this led to a speed up of about a factor 400. 

\begin{figure*}[]
	\centering
	\includegraphics[width=17cm, trim={0.0cm 0.0cm 0cm 0.0cm},clip]{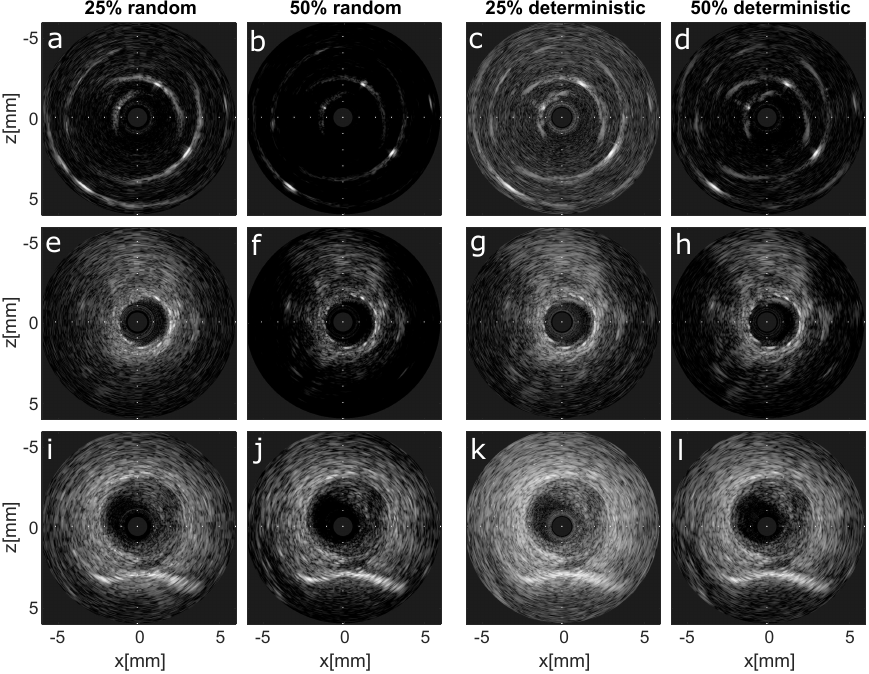}
	\caption{IVUS imaging by two-stage \acro{} from subsampled channel data at rates of 25\% and 50\% for a random and deterministic scheme. a-d) Reconstructions of simulated point scatterers, e-h) Reconstructions of an \textit{ex-vivo} arterial stent and, i-l) Reconstructions of an \textit{in-vivo} coronary artery.}
	\label{fig:subsample}
\end{figure*}

Finally, we evaluate \acro{} on two subsampling schemes. During initial experiments we observed that \acro{} was not able to adequately reconstruct images from such acquisitions directly. We therefore designed a two-stage variant of the original \acro{} architecture, which first performs beamforming on the received signals belonging to each single transmit separately, and subsequently performs weighted compounding across transmits with a second neural network. In doing so, we create a network that is twice as deep as the single stage network, but with a lower amount of trainable parameters and computational cost due to the absence of interconnections between signals belonging to different transmits in the first beamforming stage. The resulting reduction in degrees of freedom further constrains the network to the desired beamforming task, and in practice aided the reconstruction from undersampled array data. The beamformed images using both subsets (random and deterministic) with 50\% and 25\% of the transmit-receive pairs active are given in Fig.~\ref{fig:subsample}.

\section{Discussion}
\noindent For both PW and SA imaging, \acro{} yields high-contrast, high-resolution imaging that is qualitatively comparable to the MV target with significantly less clutter than DAS. Quantitatively, we see that all adaptive techniques (iMAP2, EBMV and \acro{}) display an increase in CNR and resolution compared to DAS, where \acro{} even manages to outperform its train target (EBMV) on the latter, likely due to its ability of incorporating a generalizable prior in the beamforming process by averaging statistics of the training data. 

Additionally, for IVUS imaging, adaptive beamforming yields a strong suppression of typical ringdown artifacts caused by residual vibrations after transmission. \acro{} learns this behavior from the MV beamformer, in which these artifacts, caused by high-intensity yet mostly incoherent signal components, are minimized by optimizing the weights such that the signals destructively interfere with each other. Strong scatterers that were previously obscured by the ringdown become visible due to the highly correlated nature of their echoes. 

After training the proposed network on subsampled data we again see that for both the random and deterministic strategies, \acro{} is capable of reconstructing high-contrast images. However when comparing to the full-array reconstructions, some noticable artifacts are visible, especially at 25\% channel subsampling. We see that these artifacts predominantly emerge in regions where mainly speckle is present. This is likely due to the (colored) noise-like properties of speckle, which are challenging to capture with only a limited amount of receiving channels. In contrast, strong point-scatterers remain clearly visible, relevant for applications such as arterial stent detection. 

When comparing the two subsampling strategies at equal subsampling rates, we observe noticable differences in the reconstructed images. In the randomly subsampled images we notice a wide sidelobe pattern. Yet, grating lobes appear less pronounced. A seemingly better suppression of noise and clutter in (mostly) empty regions leads to an overall better contrast than the deterministic method. 

\begin{figure}[t!]
	\centering
	\includegraphics[width=8.5cm, trim={0.5cm 0.1cm 0.5cm 0.2cm},clip]{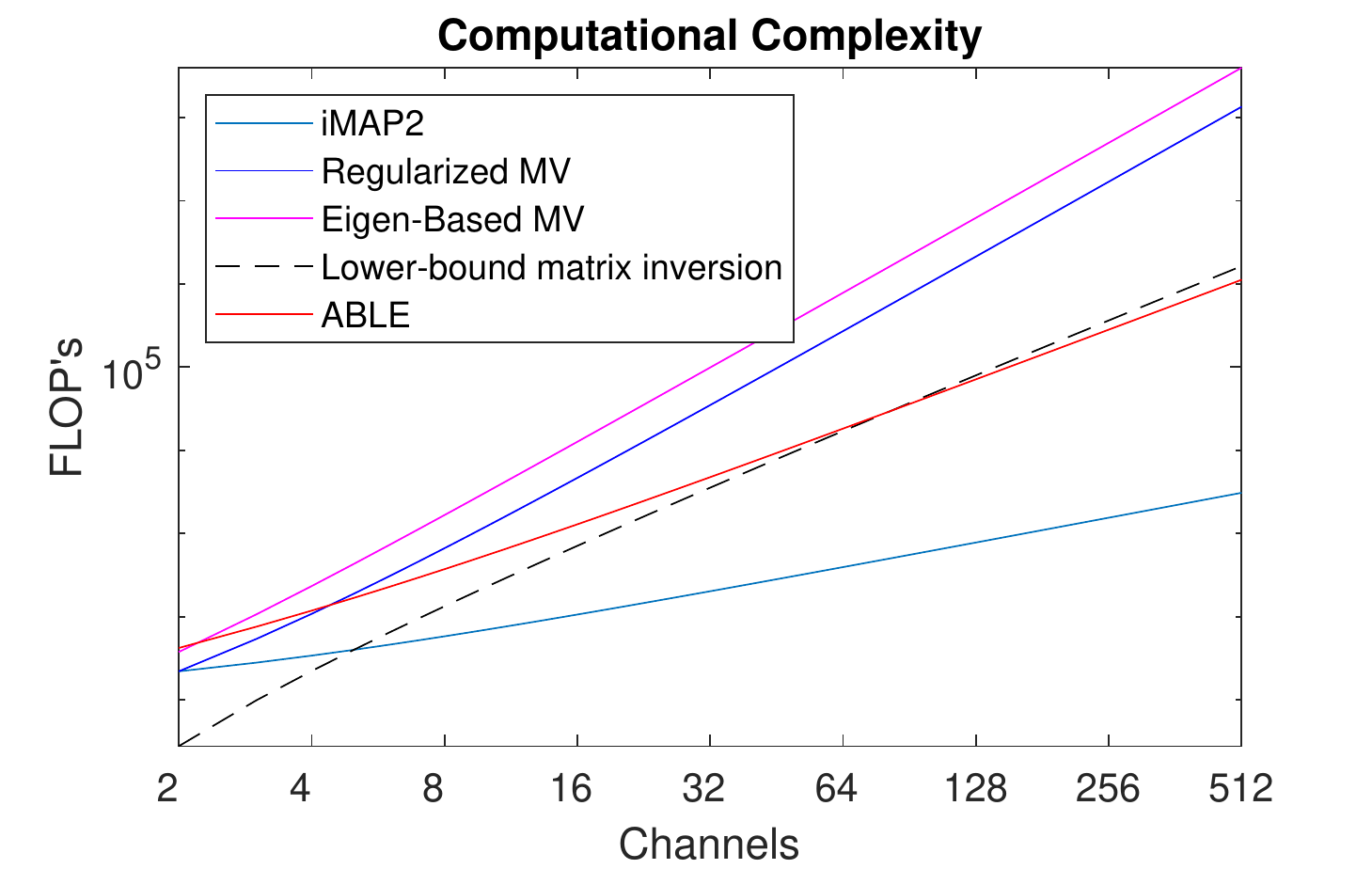}
	\caption{Number of floating-point-operations (FLOPs) for the evaluated beamforming methods as a function of the number of receiving input channels.}
	\label{fig:complexity}
\end{figure}

We here propose to leverage neural networks to learn fast approximations of adaptive algorithms. To justify this, we can asses the algorithmic complexity of these methods, which in turn is strongly correlated with their respective reconstruction times and furthermore constitutes a system-independent measure. 

In Fig. \ref{fig:complexity} the number of required FLOPs required for regularized MV, EBMV, iMAP2 and \acro{} is plotted as a function of the input array size.
Here we see that the iMAP beamformer provides a low complexity alternative compared to beamforming methods relying on the prediction of content-adaptive signal weights, however at the cost of reduced image quality. One can observe that for all array sizes, our deep learning approach provides a lower computational cost than the MV based methods relying on matrix inversion. While there are several methods to increase the efficiency of matrix inversion  \cite{Strassen:1969} \cite{Coppersmith1987}, the best proven lower bound has indicated a minimum computational cost of $N^2\log(N)$ \cite{Raz2002}, also shown in Fig. \ref{fig:complexity}. This indicates that the deep learning approach severely reduces computational burden, in particular for a large number of transducer elements. Arrays consisting of more than 64 elements are very common for ultrasound imaging, whereas 3D ultrasound systems can contain over a 1000 elements. 

The data-efficiency of the proposed network is in a large part attributed to its pixel-wise operation. At the same time this restricts the network from leveraging any knowledge about the location and context around that pixel. While this also holds for MV beamforming, it may hamper its ability to suppress artifacts such as grating lobes. This leaves ample room for future development in the direction of more complex and spatially aware deep-learning-based adaptive beamformers, in which neighboring pixels are incorporated in the adaptive parameter prediction. 

Finally, we expect the concept proposed in this work to benefit a number of other applications that require the estimation of content-adaptive signal processing parameters. One specific example is adaptive spectral Doppler \cite{Gran2009}, in which a set of optimal matched filter coefficients for high-quality spectral estimation. Due to the similarity of this problem, we expect comparable improvements in computational time and complexity. 

\pagebreak

\section{Conclusion}

\noindent In this work, we demonstrated how deep learning can be used to improve upon conventional beamforming methods. Specifically, we show that a compact and model-based architecture, which we term \acro{} (adaptive beamforming by deep learning), enables the reconstruction of high-quality ultrasound images for multiple imaging systems. Its resolution and contrast are up to par with a complex state-of-the-art adaptive beamformer, yet at a substantially lower reconstruction time and computational burden. This paves the way towards a real-time implementation of adaptive beamforming in ultrasound systems.

\bibliographystyle{ieeetran}
\bibliography{DeepBeamforming}

\end{document}